\documentclass[a4paper,10pt,twoside]{cpc-hepnp}

\usepackage{multicol}
\usepackage{graphicx}
\usepackage{booktabs}
\usepackage{amssymb,bm,mathrsfs,amscd}
\usepackage[tbtags]{amsmath}
\usepackage{lastpage}

\begin{document}

\fancyhead[co]{\footnotesize T. Matsuki et al: Derivation of a $1/r^2$ potential
term for $Q\bar Q$  in a potential model}

\footnotetext[0]{Received 14 March 2009}

\title{Derivation of a $1/r^2$ potential term for $Q\bar Q$  in a potential model}

\author{%
      T. Matsuki$^{1;1)}$\email{matsuki@tokyo-kasei.ac.jp}%
\quad Xiang Liu$^{2;2)}$\email{xiangliu@lzu.edu.cn}%
\quad T. Morii$^{3;3)}$\email{morii@radix.h.kobe-u.ac.jp}%
\quad K. Seo$^{4;4)}$\email{seo@gifu-cwc.ac.jp}%
}
\maketitle

\address{%
1~(Tokyo Kasei University,
1-18-1 Kaga, Itabashi, Tokyo 173, JAPAN)\\
2~(School of Physical Science and Technology,
Lanzhou University, Lanzhou 730000, China)\\
3~(Graduate School of Human Development and Environment,
Kobe University, Nada, Kobe 657-8501, JAPAN)\\
4~(Gifu City Women's College, 7-1 Ichinichi Ichiba Kitamachi,
Gifu 501-0192, JAPAN)\\
}

\begin{abstract}
We clarify the difference among potential models so far proposed to explain
mass spectra of heavy-light mesons via transformations of the vacuum.
Applying our idea to $Q\bar Q$ quarkonium, we obtain the extra term,
$1/r^2$ with positive coefficient, other than non-relativistic potential terms
expected for quarkonium.
\end{abstract}



\begin{multicols}{2}

\section{Introduction}

There are many articles to try to explain the heavy-light mesons since
the discovery of the so-called $D_{sJ}$, $D_{s0}^*(2317)$ and $D_{s1}'(2460)$,
by BaBar and CLEO.
Except for our model,\cite{MM97,MMS05} other relativistic potential models fail to
reproduce $D_{sJ}$ particles.
Accordingly most of the people think that these two particles are exceptional and
should be explained by some other mechanism.

We would like to explain the difference of our relativistic potential model from others
by considering the true vacuum of heavy-light system and its transformation.

\section{True vacuum and transformation}

Let us introduce the notion of the true vacuum in heavy quark effective theory.
Theory should be expanded around the true vacuum like the Higgs model.
In HQET, it can be said that whether the model pays respect on heaviness of
the heavy quark. In other word, one should expand the theory in terms of $1/m_Q$.
If it is not, then we say it is not expanded around the true vacuum.

We would like to introduce one more fact/notion that the transformation is
different from the approximation.
When one truncates the perturbation up to a certain order of a coupling constant,
this is an approximation. In this case one may use the values of parameters the
same as before truncation or closer values as before.
One can instead use transformation on the model so that we obtain the true vacuum
closer than before.
In this case, values of parameters become quite different from those before
transformation.

\subsection{HQET in field theory (Georgi transformation)}

Heavy-light system respects the heavy quark symmetry whose idea is to use $1/m_Q$
as an expansion parameter.
In the language of the field theory, it is equivalent to have a propagator which
does not propagate in the configuration but only in the time direction.
This is given by
\begin{eqnarray*}
  {\psi _v} = \exp ( - imv\cdot x)\psi,
\end{eqnarray*}
which depends on the heavy quark velocity.\cite{G90} This is a little bit modified so
that at the lowest order it can give positive energy wave function.
Here $x$ and $v$ are four-component coordinate and heavy quark classical velocity,
respectively.
The Lagrangian becomes after the transformation:
\begin{eqnarray*}
  && L = \bar \psi \left( {i - m}{\not \partial} \right)\psi  =
  {\bar \psi _v}iv\cdot\partial {\psi _v}, \\
  && \frac{{1 + \not v}}{2}{\psi _v} = {\psi _v}.
\end{eqnarray*}
This is a unitary transformation so that the probability is conserved.

\subsection{HQET in potential model}

In order to construct the heavy quark effective theory in a potential model,
we assume as in the field theory that the dominant wave function of a $Q\bar q$
bound state should include a positive-energy heavy quark state $Q_+$.
This is because at the lowest order a heavy quark becomes a static color source,
a heavy quark becomes still, and hence it can be decomposed into positive and
negative energy states.
This is realized by adopting the Foldy-Tani-Wouthuysen transformation.\cite{FWT}
The FTW has a feature that it makes a free kinetic term diagonal.
\begin{eqnarray*}
{Q^\dag }\left( {i\vec p\cdot\vec \alpha  + m_Q\beta } \right)Q \to
{Q^\dag }_{FTW}\sqrt {{m_Q^2} + {p^2}} \beta {Q_{FTW}}
\end{eqnarray*}
The explicit form of this transformation is given by
\begin{eqnarray}
  \psi  = \exp (W\,\vec {\hat p}\cdot\vec \gamma ){\psi_{FTW}},\quad
  \tan W = p/(m_Q + E), \label{eqFTW}
\end{eqnarray}
which is a unitary transformation. This gives, as expected,
\begin{eqnarray*}
  m_Q(1 - \beta ){\psi _{FTW}} = 0\quad{\rm{or}}\quad\frac{{1 + \beta }}
  {2}{\psi _{FTW}} = {\psi _{FTW}}
\end{eqnarray*}

\subsection{Lowest FTW transformation}

Let us write down the lowest non-trivial FTW transformation that gives
the second order terms in $1/m_Q$ for the effective Hamiltonian, which are the
same as those we have used in 1997.\cite{MM97}
\begin{eqnarray}
  \psi  = \left( {\begin{array}{*{20}{c}}
   1 & { - \frac{{\vec \sigma \vec p}}{{2m_Q}}}  \\
   {\frac{{\vec \sigma \vec p}}{{2m_Q}}} & 1  \\
 \end{array} } \right){\psi _{LFTW}},\quad \tan W = \dfrac{p}{2m_Q}, \label{eqLFTW}
\end{eqnarray}
which gives
\begin{eqnarray*}
  &&{U_{LFTW}}( - \vec p)\left( {\vec \alpha \cdot\vec p + \beta m_Q} \right)
  U_{LFTW}^{ - 1}(\vec p) \\
  && \quad = \beta \left( {m_Q + \frac{{{p^2}}}
  {{2m_Q}}} \right) - \frac{{{p^2}}}
  {{{m_Q^2}}}\vec \alpha \cdot\vec p + O({(1/m_Q)^3}).
\end{eqnarray*}
Compared with Eq. (\ref{eqFTW}), the transformation given by Eq. (\ref{eqLFTW})
becomes very simple four by four matrix.

\subsection{Free fermion field transformation}

The former consideration naturally extends to another transformation which is
adopted by many people. That is, many people use, at the lowest order in $1/m_Q$,
the positive-energy heavy quark field is contained in the Hamiltonian.
This transformation is given by,
\begin{eqnarray}
 \psi  &=& \left( {\begin{array}{*{20}{c}}
   1 & 0  \\
   {\frac{{\vec \sigma \vec p}}
 {{2m_Q}}} & 0  \\

 \end{array} } \right){\psi _{FFF}} \nonumber \\
 &=& \left( {\begin{array}{*{20}{c}}
   1 & { - \frac{{\vec \sigma \vec p}}
{{2m_Q}}}  \\
   {\frac{{\vec \sigma \vec p}}
{{2m_Q}}} & 1  \\

 \end{array} } \right)\frac{{1 + \beta }}
{2}{\psi _{FFF}} . \label{eqFFF}
\end{eqnarray}
This is as can be seen not a unitary transformation because of
a positive-energy projection operator, $(1+\beta)/2$.
This transformation is intentionally or not intentionally adopted by, e.g.,
Refs. \cite{MMKM88}, \cite{ZOR93}, \cite{EGF98}, etc.
This transformation only uses a positive energy component of the heavy quark
so that negative energy contribution is neglected in calculations.

\subsection{Summary of transformations}

We give in Table \ref{tab1} which paper adopts which transformation, whether
it respects HQET (true vacuum) or not, values of $D_{sJ}$.

\begin{center}
\tabcaption{ \label{tab1}  Comparison of transformations.}
\footnotesize
\begin{tabular*}{80mm}{c@{\extracolsep{\fill}}ccc}
\toprule Ref. & TF   & TV  & $D_{sJ}$ \\
\hline
\cite{GK91} & No & No & $2.48, ~2.55$ \\
\cite{MMS05} & FTW & Yes & $2.325, ~2.467$ \\
\cite{MMKM88} & FFF & Yes & $2.525, ~2.593$ \\
\cite{ZOR93} & FFF & Yes & $2.38, ~2.51$ \\
\cite{EGF98} & FFF & Yes & $2.463, ~2.535$ \\
\bottomrule
\end{tabular*}
\end{center}

In Table \ref{tab1}, Ref. means reference number, TF which transformation is used,
TV true vacuum is used or not, and $D_{sJ}$ their mass values. The observed mass
values of $D_{sJ}$ are given by $2.317$, $2.460$ GeV.

\section{Application to $Q\bar Q$}

In this section, we apply the Foldy-Tani-Wouthuysen transformation on both
two heavy quarks of quarkonium $Q\bar Q$.

To obtain the eigenvalue equation in the first order of $1/m_Q$, we introduce the Foldy-Tani-Wouthuysen
transformation for both of the heavy quarks $Q_1$ and $\bar Q_2$ in the Hamiltonian.
To make the eigenvalue equation compact and convenient, we apply the charge conjugate transformation
on the anti-heavy quark $\bar Q_2$ so that one can multiply the anti-heavy quark ($\bar Q_2$) operators from right with the wave
function and heavy quark ($Q_1$) operators from left. These operators are unitary and are explicitly given by
\begin{equation}
  U_{FTW\;i} \equiv \exp\left(W_i \vec \gamma_{Q_i} \cdot \vec t_i\right), \;
  U_c \equiv {U_{\bar Q_2}}_c=i\gamma_{\bar Q_2}^0\;\gamma_{\bar Q_2}^2,
\end{equation}
where
\begin{eqnarray}
  \tan W &=& \frac{p}{m+E}, \quad \vec t = \frac{\vec p}{p}, \;
  \left(~~p = \left|\vec p \right|~~\right) .
\end{eqnarray}
Using these definitions, the transformation is given by
\begin{eqnarray}
  &&\left(U_{total} H U_{total}^{-1}\right) \otimes \left(U_{total} \psi\right) =
  E U_{total} \psi, \nonumber \\
  && U_{total} = U_c U_{FTW\;2} U_{FTW\;1} .
\end{eqnarray}
Expanding $U_{total} H U_{total}^{-1}$, $U_{total} \psi$, and $E$ in $1/m_1$ and $1/m_2$,
we obtain a series of equations in each order of $1/m_i$.

In the case of $Q_1=Q_2$, i.e., $m_1=m_2=m$, we obtain the following non-trivial eigenvalue 
equation in the lowest order of $1/m$.
\begin{eqnarray}
  && \frac{p^2}{m}\psi_{0+1}^{s~+,+}+\left[S(r)+V(r)+\frac{9V^2}{4m} \right] \psi_{0+1}^{s~+,+} \nonumber \\
  && \quad  = E_{0+1} \psi_{0+1}^{s~+,+} ~, \label{psis} \\
  && \frac{p^2}{m}\psi_{0+1}^{v~+,+}+\left[S(r)+V(r)+\frac{V^2}{4m} \right] \psi_{0+1}^{v~+,+} \nonumber \\
  && \quad = E_{0+1} \psi_{0+1}^{v~+,+} ~, \label{psiv}
\end{eqnarray}
where the subscript "0+1" means that the non-trivial equation is obtained by
adding 0-th and 1st order
equations in $1/m$ and $E_{0+1}=E_0+E_1$. The superscripts $s$ and $v$ mean
that an $i$-th order wave function is decomposed as follows:
\begin{equation}
  \psi_i = \left( {\begin{array}{*{20}c}
     {\psi _i^{ + , - } } & {\psi _i^{ + , + } }  \\
     {\psi _i^{ - , - } } & {\psi _i^{ - , + } }  \\
  \end{array}} \right),\quad
  \psi_i^{a, b} = \psi_i^{s~a, b} + (\vec n\cdot\vec \sigma)\;
  \psi_i^{v~a, b},
\end{equation}
with $\vec n = {\vec r}/{r}$. Here we have denoted without confusion the transformed
wave function,
$\left(U_{total} \psi\right)_i$, as $\psi_i$.
For $a=\pm$, they correspond to positive and negative energy components of a heavy
quark and for $b=\pm$, they correspond to those of an anti-heavy quark.
As one can see from Eqs. (\ref{psis}, \ref{psiv}) and $V^2\propto 1/r^2$, the dominant
potential near the
origin is given by $1/r^2$ with a positive coefficient in the first order of $1/m$.
Hence one should include this term when solving Eqs. (\ref{psis}, \ref{psiv})
at the lowest non-trivial order differential equation.
Contrary to the lattice results,\cite{KKW06}
we have no problem to solve these equations because of positive coefficients for
$V^2\propto 1/r^2$ terms. When they are negative, one faces a critical value for
this coefficient below which one can not obtain a solution.\cite{KKW06,Landau}

\end{multicols}

\vspace{-2mm}
\centerline{\rule{80mm}{0.1pt}}
\vspace{2mm}

\begin{multicols}{2}

\def\Journal#1#2#3#4{{#1}, #4, {\bf #2}: #3}
\def\NIM{Nucl. Instrum. Methods}
\def\NIMA{Nucl. Instrum. Methods A}
\def\NPB{Nucl. Phys. B}
\def\PLB{Phys. Lett. B}
\def\PRL{Phys. Rev. Lett.}
\def\PRD{Phys. Rev. D}
\def\ZPC{Z. Phys. C}
\def\EPJ{Eur. Phys. J. C}
\def\PR{Phys. Rept.}
\def\IJM{Int. J. Mod. Phys. A}
\def\PTP{Prog. Theor. Phys.}

\end{multicols}

\clearpage

\end{document}